\documentclass[aps,prl,twocolumn,showpacs,superscriptaddress,floatfix,10pt]{revtex4-1}
\usepackage{amsmath,amssymb,amsfonts,stmaryrd,wasysym,graphicx,multirow,color,textcomp,subfigure}
\usepackage{url}
\usepackage[colorlinks=true,citecolor=blue,urlcolor=blue]{hyperref}
\usepackage{romannum}

\usepackage{color,soul} 
\usepackage{bm}
\usepackage[normalem]{ulem}

\hypersetup{colorlinks=true, urlcolor=blue, citecolor=cyan, pdfborder={0 0 0}}

\normalfont
\def\be{\begin{equation}}
\def\ee{\end{equation}}
\def\bea{\begin{eqnarray}}
\def\eea{\end{eqnarray}}

\newcommand{\cyan}[1]{{\color{cyan}{\em {#1}}}} 
\newcommand{\fig}[1]{Fig.\,\ref{#1}}
\newcommand{\figs}[1]{Figs.\,\ref{#1}}
\newcommand{\figr}[1]{Figure\,\ref{#1}}

\newcommand{\vv}[1]{{\boldsymbol #1}} 

\begin{document}
\title{Higher-Order Topological Insulator in Twisted Bilayer Graphene}
\author{Moon Jip Park}
\affiliation{Department of Physics, KAIST, Daejeon 34141, Republic of Korea}
\author{Youngkuk Kim}
\affiliation{Department of Physics, Sungkyunkwan University, Suwon 16419, Korea}
\author{Gil Young Cho}
\affiliation{Department of Physics, Pohang University of Science and Technology (POSTECH), Pohang 37673, Republic of Korea}
\author{SungBin Lee}
\affiliation{Department of Physics, KAIST, Daejeon 34141, Republic of Korea}

\begin{abstract}
Higher-order topological insulators are newly proposed topological phases of matter, whose bulk topology manifests as localized modes at two- or higher-dimensional lower boundaries. In this work, we propose the twisted bilayer graphenes with large angles as higher-order topological insulators, hosting topological corner charges. At large commensurate angles, the intervalley scattering opens up the bulk gap and the corner states occur at half filling. Based on both first-principles calculations and analytic analysis, we show the striking results that the emergence of the corner states do not depend on the choice of the specific angles as long as the underlying symmetries are intact. 
Our results show that the twisted bilayer graphene can serve as a robust candidate material of two-dimensional higher-order topological insulator. 
\end{abstract}
\maketitle
\cyan{Introduction} -  
Generalizing the concept of topology in diverse systems has been one of the most important topics in condensed matter physics\cite{RevModPhys.83.1057,RevModPhys.82.3045}. One recently proposed interesting path of extending the knowledge of the topological phases of matter is to consider higher-order generalization\cite{Benalcazar61}. That is to say, the band topology of $d$-dimensional insulator manifests as non-trivial $d-2$ or lower-dimensional boundary states, which is known as the higher-order topological insulator(HOTI)\cite{PhysRevB.96.245115,PhysRevLett.119.246401,PhysRevLett.119.246402,Schindlereaat0346,PhysRevLett.120.026801,Serra-Garcia2018,Peterson2018,Imhof2018,PhysRevB.97.205136,PhysRevB.97.205135,PhysRevX.8.031070,PhysRevB.97.241405,PhysRevLett.121.096803,PhysRevB.98.045125,Xue2019,PhysRevB.98.081110,PhysRevX.9.011012,PhysRevLett.121.116801,PhysRevB.97.041106,PhysRevB.98.241103,PhysRevLett.122.076801,PhysRevB.98.201402,PhysRevB.99.020304,PhysRevB.99.085406,PhysRevB.98.201114,PhysRevB.98.245102,wheeler2018many,kang2018many,ono2019difficulties}. There have been several theoretical proposals of the higher-order topological insulators in both two- and three- dimensional solid state systems. For example, in three dimensions, the higher-order topological insulators harbor one-dimensional gapless modes, which is referred to as the hinge modes. Examples include a bismuth crystal\cite{Schindler2018}, SnTe, surface-modified $\text{Bi}_2\text{TeI}$, BiSe, BiTe\cite{Schindlereaat0346}, phosphorene\cite{PhysRevB.98.045125}, van der Waals multilayeres\cite{PhysRevB.98.245102}, and transition metal dichalcogenides $\text{XTe}_2$ \cite{wang2018higher}. In two dimensions,  $0$-D localized modes appear at the corners of the two-dimensional materials, known as the corner states. The proposed examples are two-dimensional phosphorene\cite{PhysRevB.98.045125}, monolayer graphdiyne\cite{lee2019higher,sheng2019two}, and effective model of the twisted bilayer graphene\cite{PhysRevX.9.021013,PhysRevLett.114.226802}. 

Although there already exist a few candidate materials for the two-dimensional HOTIs, experimental signature that unambiguously identifies such phases remains elusive to date. To this end, it is desirable to identify higher-order topological materials that are readily available and highly controllable such as bismuth and other two-dimensional materials uncovered in Ref. \cite{prep}. Here, we propose that a large angle twisted bilayer graphene is a generic higher-order topological insulator, characterized by the topologically protected corner states.  Twisted bilayer graphene has clear advantages over other candidates for the experimental detection of the corner charge: (i) In our proposal, the HOTI is realized at half filling, thus corner state occurs without any fine-tuning of chemical potential, which is distinct from other cases\cite{PhysRevX.9.021013,1807.10676}.
(ii) In graphene, ultra clean transfer techniques are available that enables isolation from undesirable substrate effects\cite{lee2017review}.
(iii) The HOTI in graphene is very unique system realized by intervalley scattering with negligible spin-orbit coupling. Thus, one can avoid any ambiguity arising from the spin-orbit coupling effect. 

\begin{figure}[t!]
	\centering\includegraphics[width=0.45\textwidth]{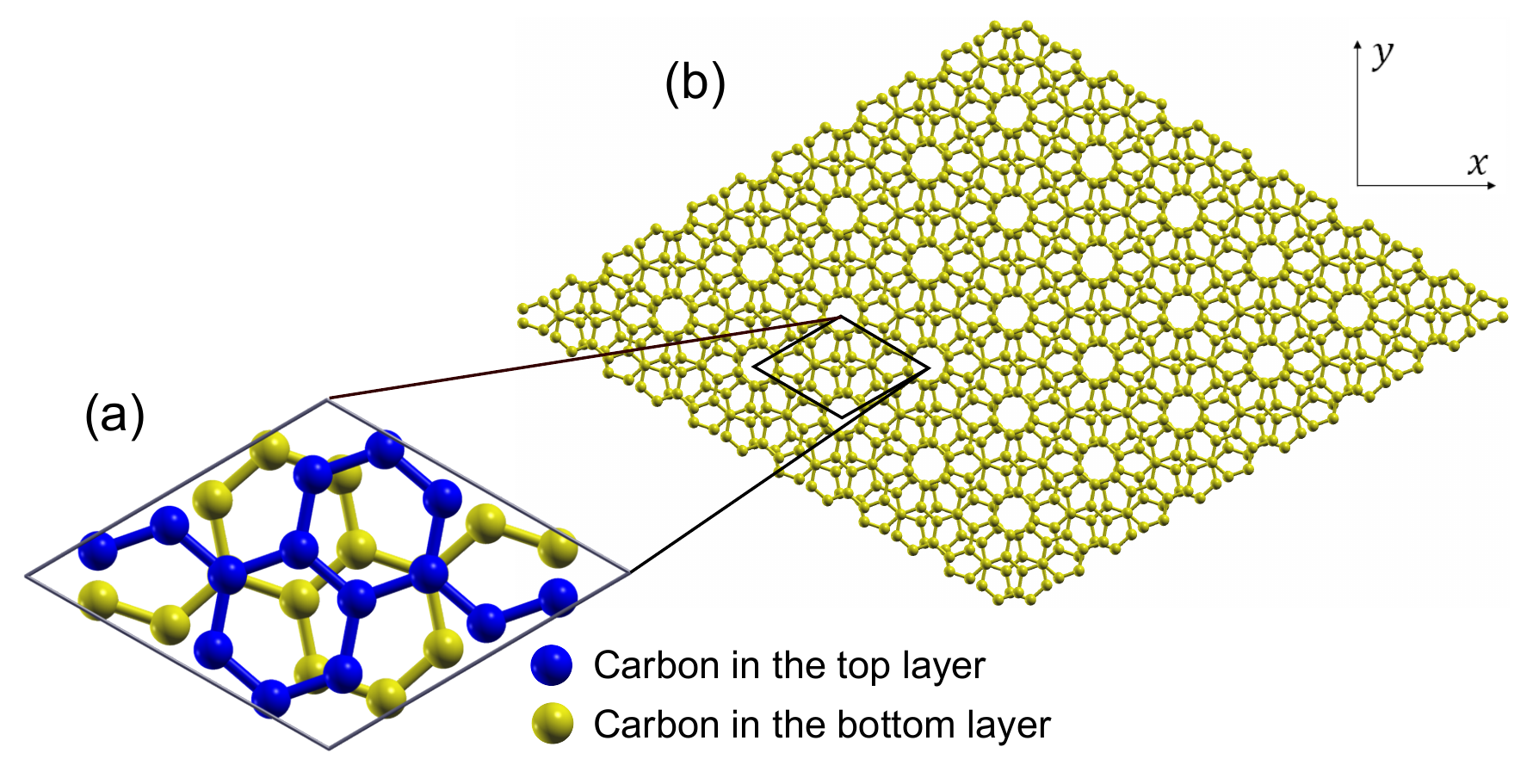}
	\caption{Atomic configurations of the twisted bilayer graphene in (a) a single Moir\'{e} unit cell and (b) a finite-sized system with the twist angle, $\theta=21.78^\circ$ ($p=1$, $q=3$). }
\label{fig:as} 
\end{figure}

In this work, for the first time, we propose that the twisted bilayer graphenes with large commensurate angles can be generically the higher order topological insulators. 
We first present the first-principles calculations of the twisted bilayer graphene at the commensurate angle, $\theta=21.78^\circ$, as a representative example. In this case, we find that the sizable gap ($\sim$9\ meV) exists which is originated from the absence of the $U(1)$ valley symmetry, $U(1)_v$. We then show that twisted bilayer graphene has non-trivial higher-order band topology, characterized by the occurrence of topological corner states with the fractional electron charge $e/2$ in the mirror symmetric corners and $C_6$ symmetric corners. In addition, we generalize our discussion by showing that the non-trivial band topology is guaranteed to exist regardless of the specific commensurate twist angle or the microscopic details of the atomic structure as long as the underlying symmetries are preserved. 

\cyan{Lattice model and Symmetries} - 
We begin our discussion by introducing the atomic structure and the associated crystalline symmetries of twisted bilayer graphene. We consider a specific set of twisted bilayer structures that belong to the hexagonal space group \# 177 (point group  $D_6$). The atomic configuration can be readily constructed by twisting AA-stacked bilayer graphene with respect to the collinear axis at the hexagonal center [See \figs{fig:as}(a) and \ref{fig:as}(b)]. The twist preserves both $C_{6z}$ and $C_{2x}$ about the out-of-plane $z$- and in-plane $x$-axes, respectively, which generate another important rotation $C_{2y}$ about the $y$-axis. The original translation symmetry is broken by the twist, but we can define the Moir\'{e} translational symmetry, depending on the twist angle. For any coprime integers $p$ and $q$,  a twist by $\theta_{p,q}\!=\!\arccos \frac{3p^2+3pq+q^2/2}{3p^2+3pq+q^2}$ results in an enlarged Moir\'{e} unit cell with the lattice constant $L \!=\! a\sqrt{\frac{3p^2+3pq+q^2}{{\rm gcd}(q,3)}}$ \cite{PhysRevB.98.085435,PhysRevLett.99.256802}, where $a$ is the original lattice constant and gcd represents the greatest common divisor.  

\cyan{Electronic energy bands} - 
In the limit where $\theta \! \lesssim \!1^\circ$ without the lattice distortions, the Moir\'{e} potential has long periodicity in real space, resulting in negligible interaction between valleys.  In this limit, including the so-called {\em magic angles} where the Fermi velocity vanishes\cite{Bistritzer12233}, each valley is decoupled and the  $U(1)$ valley symmetry $U(1)_v$ is approximately preserved and it, together with $C_{2z}\mathcal T$ symmetry, provides topological protection of four Dirac points associated with the $\mathbb{Z}_2$-quantized Berry phases $\pi$ \cite{PhysRevLett.118.156401, 1807.10676}. Here, $\mathcal T$ represents time-reversal symmetry, where $\mathcal T^2 = 1$ without spin-orbit coupling.   

However, in the large angle limit $\theta \rightarrow 30^\circ$, which is the concern of this paper, the $U(1)_v$ symmetry can be broken and the fourfold-degenerate Dirac points can split into two pairs of massive Dirac points, which here we claim to induce a HOTI phase. The intervalley coupling, and thus the gap opening between Dirac points, has a tendency to increase as the twist angle $\theta$ increases. Eventually, a sizable band gap of around 10\ meV order opens at $\theta=21.78^\circ$ ($p=1$, $q=3$)\cite{PhysRevLett.101.056803}. Focusing on this particular angle first, below we demonstrate the HOTI phase in twisted bilayer graphene using first-principles calculations.

\begin{figure}[t!]
	\centering\includegraphics[width=0.45\textwidth]{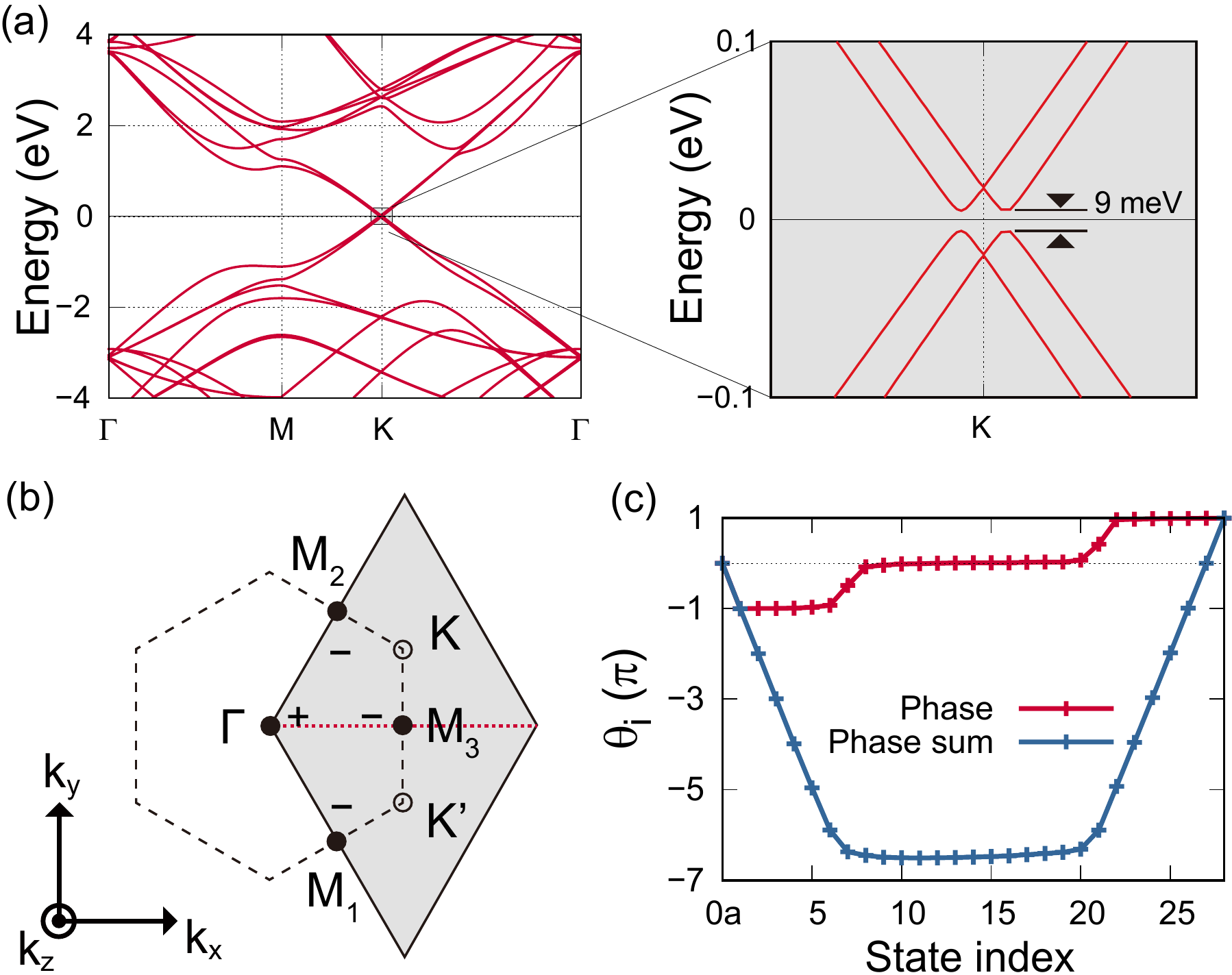}
   \caption{
   (a) DFT band structures of twisted bilayer graphene with $\theta = 21.78^\circ$. The right panel shows the magnified view of the region near $K$ indicated by a gray box in the left panel.  (b) Moir\'{e} Brillouin zone and the parity eigenvalues for half the occupied bands with negative parity. Time-reversal invariant momenta are indicated by a solid circle. A dotted red line highlights the high-symmetry $C_{2x}$-invariant line at $k_x = 0$. (c) Zak phase calculated using the mirror +1 bands along the mirror-invariant $k_x = 0$ line.  The red curve represents the $i$-th eigenphase $\theta_i$. The blue curve represents the accumulated phase up to $i$-th states, which results in $\pi$ when summed up over all the occupied mirror-$+1$ bands. The mirror winding number is calculated for the 56 occupied bands, comprising the same number of mirror-$+$ and mirror-$-$ bands ($N_{\rm occ}^+ = N_{\rm occ}^- = 28$). 
   }
\label{fig:es} 
\end{figure}

\figr{fig:es}(a) shows the density functional theory (DFT) electronic band structures of the twisted bilayer graphene with $\theta$ = 27.78$^\circ$.  The results show that the band structure features a narrow gap of $\sim$ 9\ meV, which is more or less similar to the previous result of $\sim$ 7\ meV \cite{PhysRevLett.101.056803}. While a global direct band gap appears throughout the entire Moir\'{e} BZ between the conduction and valence bands, the minimum value occurs in the vicinity of $K$ (as well as $K'$) as shown in the right panel of \fig{fig:es}(a). Using all the occupied 56 bands, we explore the higher-order band topology by directly calculating a mirror-winding number and the second Stiefel-Whitney number.

\cyan{Mirror-winding number} - 
Whereas we find that the Zak phase along any closed loop is trivial, a finer topological classification can be made along the mirror invariant line utilizing the presence of the mirror symmetry. The BZ has a mirror-invariant line $\Gamma$-$M_3$-$\Gamma$ preserving $C_{2x}$ symmetry [See \fig{fig:es}(b)]. Along this mirror invariant line, we can decompose the Hamiltonian into two distinct sub-sectors characterized by the mirror eigenvalues $\pm 1$. Then, one can define the $\mathbb{Z}_2$ mirror winding number \cite{RevModPhys.88.035005,PhysRevB.88.075142} as the mirror-resolved Zak phase, $\nu_\pm$, calculated along $\Gamma$-$M_3$-$\Gamma$ line for $C_{2x}=\pm 1$ subsector. The $\mathbb{Z}_2$ mirror winding number  $\nu$ can be evaluated by 
\begin{align}
\nu = \nu_+ = \nu_- ~({\rm mod} ~ 2),
\end{align}
where
\begin{align}
 \nu_\pm = \frac{1}{i\pi}\log\det[\mathcal{U}_\pm(\Gamma,M_3)\mathcal{U}_\pm(M_3,\Gamma)].
 \end{align}
Here,   $\mathcal{U}_\pm(\vv{k}_1,\vv{k}_2) \equiv  P \exp[i\int_{\vv{k}_1}^{\vv{k}_2}\boldsymbol {\mathcal{A}_\pm}(\vv{k})\cdot d\vv{k} ]
= \tilde{P}_\pm(\vv{k}_1)[\prod_\vv{k} \tilde{P}_\pm(\vv{k})] \tilde{P}_\pm(\vv{k}_2)$, $\mathcal{A}_\pm$ is the non-Abelian Berry connection evaluated from the $\pm$-sector, respectively. $P$ indicates the path ordering, and $\tilde{P}_\pm$ is the projection operator to the occupied  mirror $\pm$-subspace. Using all the occupied DFT bands, we calculate the eigenphase $\theta_{\pm,i} \equiv {\rm Im}\log (u_{\pm,i})$, where $u_{\pm,i}$ is the $i$-th eigenvalue of  the Wilson loop matrix $\mathcal{U}_\pm(\Gamma,M_3)\mathcal{U}_\pm(M_3,\Gamma)$. The results, presented in \fig{fig:es}(c), show that $\nu = \nu_\pm = +1$, confirming the non-trivial band topology.  

\cyan{Stiefel-Whitney number} - 
The physical manifestation of the non-trivial winding number $\nu = 1$  is the emergence of the topological corner states as we discuss later.  Such corner states can be further established by the second Stiefel-Whitney number \cite{PhysRevLett.121.106403}. The twisted bilayer graphene preserves {the two-dimensional inversion symmetry} $C_{2z}$, whose product with time-reversal symmetries $C_{2z} \mathcal {T}$ imposes that the Hamiltonian is real symmetric matrix, thus being characterized by the Stiefel-Whitney classes. While the first Stiefel-Whitney number $\omega_1$, which is equivalent to the Zak phase, is turned out to be trivial,  the non-trivial second $\mathbb{Z}_2$ Stiefel-Whitney number $\omega_2$ is expected from our DFT calculations. As proposed by Ahn {\em et al.} \cite{PhysRevLett.121.106403, 1904.00336}, we evaluate $\omega_2$ from the parity eigenvalues of occupied bands at time-reversal invariant momenta (TRIM) $\Gamma_i \in \{ \Gamma, M_1, M_2, M_3\}$ using
\begin{align}
(-1)^{\omega_2} =  \prod_{\Gamma_i\in{\rm TRIM} } (-1)^{[N_{\rm occ}^-(\Gamma_i)/2]},
\label{eq:SW}
\end{align}
where $N_{\rm occ}^-(\Gamma_i)$ is the number of occupied bands with an odd parity at $\Gamma_i$. 
\figr{fig:es}(c) shows the calculated parity eigenvalues at TRIM. We find that $\Gamma$ ($M_{1,2,3}$) has 24 (30) odd bands out of the 56 occupied bands. Therefore the parities at $(\Gamma,M_1,M_2,M_3) = (+,-,-,-)$, leading to $\omega_2 = 1$.

\begin{figure}[t!]
	\centering\includegraphics[width=0.45\textwidth]{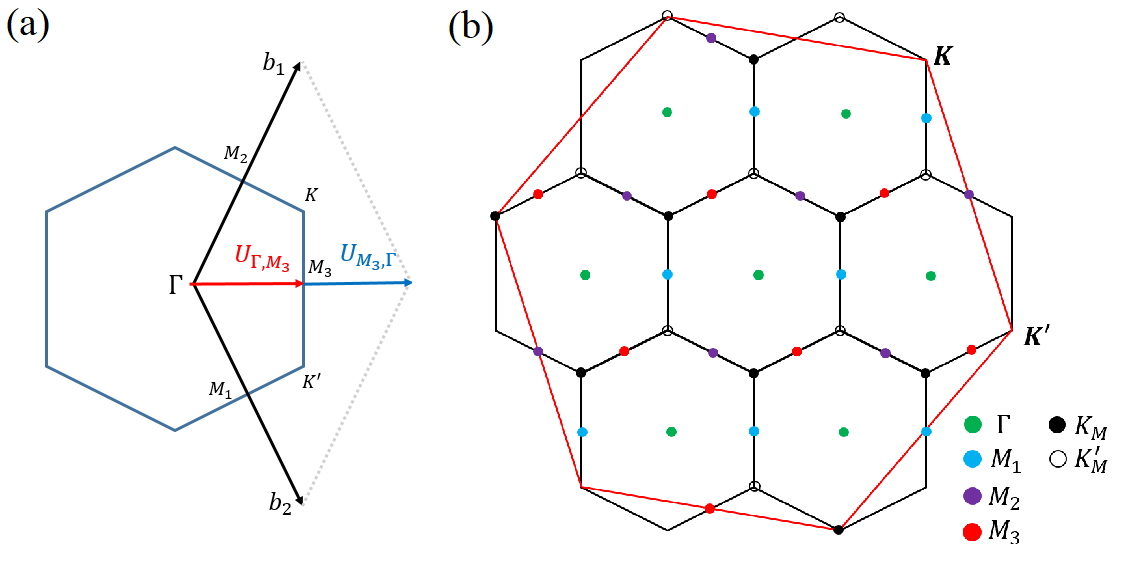}
	\caption{(a) Schematic figure representing the Wilson loop along the mirror invariant line. We can decompose the Wilson loop into two pieces(red and blue lines). These two lines are related by the inversion symmetry. (b) Figure of monolayer BZ and Moir\'{e} BZ when $\theta=21.78^\circ$. In the monolayer BZ, the TRIM points in Moir\'{e} BZ are always replicated odd times (In this case, $7$ times.).}
	\label{fig:wilsonline} 
\end{figure}

The mirror winding number $\nu$ and the second Stiefel-Whitney number $\omega_2$ are, in general, irrelevant topological invariants that depend on different symmetries ($C_{2x}$ and $C_{2z}\mathcal{T}$, respectively).  However, in the twisted bilayer graphene,where both symmetries are present, we can formally equate these numbers. We decompose the Wilson loop into the two pieces of the Wilson lines related by the  inversion symmetry [See \fig{fig:wilsonline}(a)].  In this case, one can show the exact cancellation between the two Wilson lines up to the parity eigenvalues of $C_{2z}$ as following,

\begin{align}
\label{Eq:Wilson}
&\det[\mathcal{U}_\pm(\Gamma,M_3)\mathcal{U}_\pm(M_3,\Gamma)]
\\
\nonumber
&=\det[\mathcal{U}_\pm(\Gamma,M_3)
C_{2z}(C_{2z}^{-1}\mathcal{U}_\pm(M_3,\Gamma)C_{2z}) C_{2z}^{-1}]
\\
\nonumber
&=\prod_{n\in {\rm occ},C_{2x}=\pm1} \zeta_n(\Gamma) \zeta_n(M_3),
\end{align}
where $\zeta_n(\vv{k})$ is the eigenvalue of $C_{2z}$ at $\vv{k}$ on $n$-th band. Since $C_{3z}$ guarantees the same parity structure of the bands at $M_1$, $M_2$, and thus $\zeta_n(M_1)\zeta_n(M_2) = 1$ for any band index $n$. Therefore, we finally equate the mirror winding number $\nu = \nu_\pm$ to the second Stiefel-Whitney number  $\omega_2$,
\bea
\nonumber
(-1)^{\nu_+}
&=&
\prod_{C_{2x}=1,n\in {\rm occ}} \zeta_n(\Gamma) \zeta_n(M_3) \zeta_n(M_1) \zeta_n(M_2)
\\
&=&\prod_{\Gamma_i\in\text{TRIM}} (-1)^{N^-_{occ}(\Gamma_i)/2}\equiv (-1)^{\omega_2}.
\label{eq:SW2}
\eea
In the last line we used the absence of Zak phase (the first Stiefel-Whitney number $\omega_1 = 1$) along $\Gamma$-$M_3$-$\Gamma$, which enforces the same parity between the mirror-even and mirror-odd sectors \footnote{See supplementary material at http://XXXXXXX.XXXX.XXX for the detailed proof of Eq.\,\eqref{eq:SW2}.
}.

\cyan{Topological corner states}-
\begin{figure}[t!]
	\centering\includegraphics[width=0.45\textwidth]{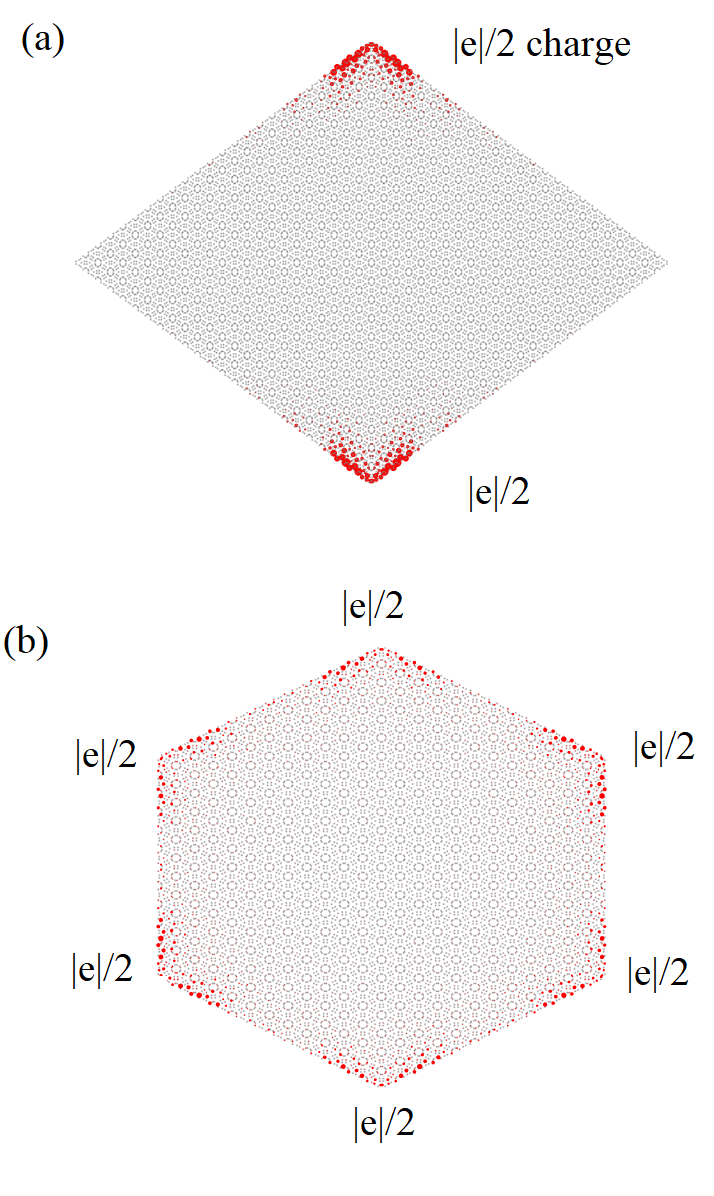}
	\caption{The topological corner states derived from the tight binding model with (a) mirror symmetric and (b) $C_6$ symmetric open boundary conditions. The size of the red circles indicate the wave function amplitude. We find $\frac{e}{2}$ localized states on the corners in both boundary conditions. We have used the same tight-binding parameters as ref. \cite{PhysRevB.87.205404}}.
	\label{fig:corner}
\end{figure}
After presenting the non-trivial topology of the bulk using the first-principles calculations, we utilize the tight-binding model\cite{PhysRevB.87.205404} to demonstrate the topological corner states in a large size twisted bilayer graphene (more than $20,000$ atoms). We take the open boundary condition preserving the mirror symmetry. \figr{fig:corner}(a) shows the tight-binding model calculations of the localized fractional e/2 charge existing at the mirror symmetric corners. In addition, one can also consider the boundary termination with $C_6$ symmetry. In such case, we can define $\mathbb{Z}_6$ valued $C_6$ quadrupole moment. This value is explicitly given as \cite{benalcazar2018quantization},
\begin{gather}
Q^{(6)}=\frac{e}{2}w_2+\frac{e}{6}(\#C_3 ) \enskip (\text{mod} \enskip e),
\label{Eq:q6}
\end{gather}
where $\#C_3$ counts the difference in the number of the occupied bands with $\hat{C}_3$ eigenvalue $1$ between $K$ and $\Gamma$. We find that the second term in Eq. \eqref{Eq:q6} vanishes. Accordingly, $e/2$ localized corner charge occurs at each $C_6$ symmetric corners. \fig{fig:corner} shows such localized corner states derived from the tight binding model.

\cyan{Generalization to an arbitrary twist angle} - 
It is important to note that the Stiefel-Whitney number, as well as the mirror winding number, is determined by the eigenstates at TRIM, whereas the low-energy physics of the twisted bilayer graphene is described in the vicinity $K$ and $K'$ points in the Moir\'{e} BZ. Therefore, the interlayer coupling does not alter the occupations of the TRIM points as compared to the non-interacting monolayer graphenes. As a consequence, the eigenvalues of $C_{2z}$ and the Stiefel-Whitney number remains robust irrespective on the microscopic details of the interlayer coupling as long as the associated symmetries are preserved. 
Therefore, we can extend our analysis to arbitrary commensurate angles without the microscopic band structure calculations, assuming that all the low energy physics is described near $K$ and $K'$ points. Using this special property, we now show that the twisted bilayer graphene in all other commensurate angles (arbitrary $p,q$) are higher-order topologically non-trivial at half-filling. To do so, we first need to count the eigenvalues of inversion symmetry for the non-interacting monolayer graphene at the TRIM points in the Moir\'{e} BZ. We first notice that $\frac{3p^2+3pq+q^2}{gcd(q,3)}\equiv 2N+1$ ($N\in \mathbb{Z}$) is always an odd integer and the area of the monolayer BZ fills $2N+1$ numbers of the Moir\'{e} BZ. For example, \fig{fig:wilsonline}(b) shows that the monolayer BZ fills $7$ times of Moir\'{e} BZ at $\theta=21.78^\circ$. In generic commensurate angles, there exist $2N+1$ occupied states at each TRIM point, say $\vv{X}$, in the Moir\'{e} BZ. In the original BZ of a monolayer, these states correspond to the occupied state of the monolayer graphene at distinct `non-TRIM' points related by the Moir\'{e} reciprocal vectors [See \fig{fig:wilsonline}(b)]:
\bea
\vv{X}_{m,n} \equiv \vv{X} + m\vv{b}_{M,1}+n\vv{b}_{M,2},
\eea
where $m,n$ are some integers. Among all the possible `independent' $2N+1$ numbers of momentum points $\vv{X}_{m,n}$, only $\vv{X}_{0,0}$ is the inversion symmetric point in the monolayer BZ. Other $2N$ distinct momenta, $\vv{X}_{m,n}$, are mapped to $\vv{X}_{-m,-n}$ under inversion symmetry. Explicitly, one can choose a gauge such that $\hat{C}_{2z}$ symmetry operator acts in the occupied eigenstates space as, $\hat{C}_{2z}|\vv{X_{m\neq 0,n\neq 0}} \rangle=|\vv{X}_{-m,-n} \rangle$. If we choose the basis of the occupied states at the TRIM points in the Moir\'{e} BZ as $(|\vv{X}_{0,0} \rangle,|\vv{X}_{0,1} \rangle, |\vv{X}_{-0,-1} \rangle,...,|\vv{X}_{m,n} \rangle, |\vv{X}_{-m,-n} \rangle)^T$, $\hat{C}_{2z}$ operator in the Moir\'{e} BZ is explicitly represented as,
\begin{gather}
\hat{C}_{2z}=\left(
\begin{array}{cc}
\zeta(\vv {X}) & 0 \\
0 & \sigma_x \otimes I_{N} \\
\end{array}
\right),
\label{eq:C2Z}
\end{gather}
where $\zeta(\vv {X})$ is the inversion eigenvalue of the occupied state at $\vv{X}$ in the monolayer graphene. Applying the eigenvalues of Eq. \eqref{eq:C2Z} to the second Stiefel-Whitney number in Eq. \eqref{eq:SW}, the contributions from $\vv{X}_{m\neq 0,n\neq 0}$ cancels out and we find that 
\bea
(-1)^{\omega_2}=\zeta(M)\zeta(\Gamma).
\eea
In the monolayer graphene, the inversion eigenvalues at $M$ point and $\Gamma$ point always differ by the sign, so the second Stiefel-Whitney number must be always non-trivial, $\omega_2=1$. As a consequence, the twisted bilayer graphene at all commensurate angles must be higher-order topologically non-trivial at half filling, as long as the gap is present at $K$ point. In principle, the topological transition to the trivial state is possible if the additional gap closing occurs at $M$ point. However, such gap closing contradicts with the previous first-principle calculations\cite{PhysRevB.81.165105}. 

\cyan {Conclusion} - 
In summary, we have studied the higher-order topological properties of the twisted bilayer graphene. When the twisted angle is large enough, the Moir\'{e} pattern becomes short ranged, the intervalley scattering plays an important role. 
It results in $U(1)_v$ symmetry breaking and the twisted bilayer graphene system opens up the gap even at charge neutrality. Based on the first principle DFT calculation, we have demonstrated the existence of gap. We have also shown that the twisted bilayer graphene naturally hosts topological corner states that are protected by the mirror winding number. It is worth emphasizing the emergence of the HOTIs in twisted bilayer graphene at any commensurate angles is very generic properties induced by intervalley scattering and exists in any commensurate twist angles as long as the system opens a gap. In addition, the corner states occur at the half filling such that it does not require any fine tuning of chemical potential. Our work provides important guidance for the search of the higher-order topological materials and paves the way for future experiments in the twisted bilayer graphenes.

\acknowledgments
We appreciate Byungmin Kang, Seoung-Hun Kang and Young-Woo Son for fruitful discussion. M.J.P. and S.B.L. are supported by the KAIST startup, National Research Foundation Grant (NRF-2017R1A2B4008097) and BK21 plus program, KAIST. Y.K. was supported from NRF Grant (No. NRF-2019R1F1A1055205). The
computational resource was provided from the Korea Institute of Science and Technology Information (KISTI).

\renewcommand{\thefigure}{S\arabic{figure}}
\setcounter{figure}{0}
\renewcommand{\theequation}{S\arabic{equation}}
\setcounter{equation}{0}

\begin{widetext}
\section{Supplementary Material for ``Higher-Order Topological Insulator in Twisted Bilayer Graphene''}
\subsection{Methods for first-principles calculations}
Our first-principles calculations were carried out based on density functional theory (DFT) as implemented in the \textsc{Quantum Espresso} package \cite{giannozzi09p395502}.  We employed 
local density approximation (LDA)  with the Perdew-Zunger type exchange-correlation energy \cite{PhysRevB.23.5048}. A norm--conserving, optimized, designed nonlocal pseudopotential is generated for describing the core electrons of carbon using the {\sc Opium} package  \cite{rappe90p1227, Ramer99p12471}.  The electronic wave functions were expanded in terms of a discrete set of plane-waves basis within the energy cutoff of 680 eV.   The 30$\times$30$\times$1 $\vv k$-points were sampled from the first Brillouin zone (BZ) using the Monkhorst-Pack scheme \cite{Monkhorst76p5188}. The atomic structures were fully relaxed within a force tolerance of 0.005 eV/\AA.  The van der Waals interaction described described based on the semiempirical dispersion-correction DFT (DFT-D) method \cite{Grimme06p1787}.

\subsection{Detailed Calculation of Mirror winding number}
In this section, we provide the detailed calculation of the mirror winding number. We start with the Wilson lines in Eq. \eqref{Eq:Wilson}. We can related the Wilson loop and the inversion eigenvalues as following,

\begin{align}
\nu_{\pm}&=\det[\mathcal{U}_\pm(\Gamma,M_3)\mathcal{U}_\pm(M_3,\Gamma')]
\\
\nonumber
&=\det[\langle u_i(\Gamma)| \mathcal{U}_\pm(\Gamma,M_3)\mathcal{U}_\pm(M_3,\Gamma') | u_j(\Gamma)\rangle]
\\
\nonumber
&=\det[\langle u_i(\Gamma)| \mathcal{U}_\pm(\Gamma,M_3) 
\hat{C}_{2z} (\hat{C}_{2z}
\mathcal{U}_\pm(M_3,\Gamma')\hat{C}_{2z}) \hat{C}_{2z} | u_j(\Gamma)\rangle]
\\
\nonumber
&=
\det[\sum_{a,b,c\in occ.}
\langle u_{i}(\Gamma) |\mathcal{U}_{\pm} (\Gamma , M_3)|u_{a}(M_3)\rangle
\langle u_{a}(M_3) |\hat{C}_{2z}|u_{b}(M_3)\rangle
\langle u_{b}(M_3) |\mathcal{U}_{\pm}( M_3 ,\Gamma)|u_{c}(\Gamma)\rangle
\langle u_{c}(\Gamma) |\hat{C}_{2z}|u_{j}(\Gamma)\rangle
]
\\
\nonumber
&=\det[\sum_{a\in occ.} \langle u_i(\Gamma)| \hat{C}_{2z} | u_a(\Gamma)\rangle
\langle u_a(M_3)|\hat{C}_{2z} | u_j(M_3)\rangle]
\\
\nonumber
&=\prod_{n\in {\rm occ},C_{2x}=\pm1} \zeta_n(\Gamma) \zeta_n(M_3),
\end{align}
where $u_n(\mathbf{k})$ represents the eigenstate of $n$-th band at the momentum $\mathbf{k}$. We also use the notation $\Gamma'\equiv\Gamma+\mathbf{b_1}+\mathbf{b_2}$ to avoid the confusion between $\Gamma-M_3$ and $M_3-\Gamma'$ Wilson lines. Since $\sum_{n\in occ}\zeta_n (M_1)\zeta_n(M_2)=1$ due to $C_{3}$ symmetry, we can rewrite the winding number as,
\begin{align}
\nu_{\pm}=\prod_{n\in {\rm occ},C_{2x}=\pm1} \zeta_n(\Gamma)\zeta_n (M_1)\zeta_n(M_2) \zeta_n(M_3).
\end{align}
Since $\nu_{+}+\nu_{-}=0 ~({\rm mod} ~ 2)$, we formally equate the mirror winding number to the second Stiefel-Whitney number as follows.
\begin{align}
\nu_{\pm}=\prod_{\Gamma_i\in\text{TRIM}} (-1)^{N^-_{occ}(\Gamma_i)/2}\equiv (-1)^{\omega_2}.
\end{align}
\end{widetext}

\bibliography{reference}

\end{document}